\documentclass[prl,twocolumn,showpacs, showkeywords,amsmath,amssymb,superscriptaddress,nofootinbib]{revtex4-1}

\usepackage{amssymb}
\usepackage{amsmath}
\usepackage{epsfig}
\usepackage{hyperref}
\usepackage{breakurl}
\usepackage{xcolor}

\usepackage{graphicx, subfigure, placeins, float}

\usepackage{extarrows}

\usepackage{multirow} 
\usepackage{longtable} 

\usepackage{romannum}

\newcommand{\itbf}[1]{\textbf{\textit{#1}}}

\begin{document}

\title{Color-Kinematics Duality and Dual Conformal Symmetry \\ for A Four-loop Form Factor in ${\cal N}=4$ SYM }
\author{Guanda Lin}
\email{linguandak@pku.edu.cn}
\affiliation{CAS Key Laboratory of Theoretical Physics, Institute of Theoretical Physics, Chinese Academy of Sciences,  Beijing, 100190, China}
\author{Gang Yang}
\email{yangg@itp.ac.cn}
\affiliation{CAS Key Laboratory of Theoretical Physics, Institute of Theoretical Physics, Chinese Academy of Sciences,  Beijing, 100190, China}
\affiliation{School of Physical Sciences, University of Chinese Academy of Sciences, Beijing 100049, China}
\affiliation{School of Fundamental Physics and Mathematical Sciences, Hangzhou Institute for Advanced Study, UCAS, Hangzhou 310024, China}
\affiliation{International Centre for Theoretical Physics Asia-Pacific, Beijing/Hangzhou, 310024, China}
\author{Siyuan Zhang\vspace{2mm}}
\email{zhangsiyuan@itp.ac.cn}
\affiliation{CAS Key Laboratory of Theoretical Physics, Institute of Theoretical Physics, Chinese Academy of Sciences,  Beijing, 100190, China}
\affiliation{School of Physical Sciences, University of Chinese Academy of Sciences, Beijing 100049, China}

\begin{abstract}

We obtain the integrand of full-color four-loop three-point form factor of the stress-tensor supermultiplet in ${\cal N}=4$ SYM, based on the color-kinematics (CK) duality and generalized unitarity method. 
Our result not only manifests all dual Jacobi relations via CK duality but also contains 133 free parameters.
This suggests the constructibility of the form factor at even higher loops via CK duality.
We also find that the planar form factor has a hidden dual conformal symmetry in the lightlike limit of the operator momentum, which is checked up to four loops. 
\end{abstract}



\maketitle

\section{Introduction}
\label{sec:introduction}

\noindent
Recent years have witnessed tremendous progress in our understanding of scattering amplitudes, for which the ${\cal N}=4$ super Yang-Mills (SYM) theory has been an important testing ground, see \emph{e.g.}~\cite{Elvang:2013cua,Henn:2014yza,Dixon:2013uaa, Alday:2008yw, 2011JPhA44a0101R} for an introduction.
These developments not only provide new computational methods but also reveal abundant hidden structures and symmetries.
For example, in the planar ${\cal N}=4$ SYM,  the dual conformal symmetry was discovered for amplitudes and is related to the duality between amplitudes and lightlike Wilson loops \cite{Drummond:2006rz, Alday:2007hr, Drummond:2007aua, Brandhuber:2007yx, Drummond:2007au}. 
Beyond the planar limit, an intriguing duality between color and kinematics of amplitudes was discovered in  \cite{Bern:2008qj, Bern:2010ue}, which has significantly enhanced our understanding of the non-planar sector of gauge theories and led to new high-loop results in both (super)Yang-Mills and (super)gravity theories; an extensive review can be found in \cite{Bern:2019prr}.

Inspired by the great success in amplitudes, it is natural to also dig into other important quantities in quantum field theory, such as local operators and correlation functions.
In this respect, form factors,  as a class of matrix elements involving both asymptotic on-shell states and local operators, have drawn increasing attention in recent years; see \cite{Yang:2019vag} for a recent review.
In this paper we will consider the three-point form factor of the stress-tensor supermultiplet ${\cal T}$ in $\mathcal{N}=4$ SYM \cite{Brandhuber:2012vm,Lin:2021kht,Dixon:2020bbt,Dixon:2021tdw,Dixon:2022rse}:
\begin{equation}
\itbf{F}_{3}(1, 2, 3;q) = \int d^{D} x e^{-i q \cdot x}\langle \Phi(p_1)\Phi(p_2) \Phi(p_3)|\mathcal{T}(x)| 0\rangle,
\end{equation}
where $q=\sum_{i=1}^3 p_i$ is the off-shell momentum carried by the operator. 
From the phenomenological point of view, this form factor is also interesting as it is expected to provide the maximally transcendental part of the Higgs-to-three-gluon amplitude in QCD \cite{Kotikov:2002ab,Kotikov:2004er, Brandhuber:2012vm, Gehrmann:2011aa}.
Our targets are to construct this form factor with full color-dependence at the unprecedented fourth loop order, and to study its structure of both the color-kinematics duality and the dual conformal symmetry up to this order.

The color-kinematics (CK) duality \cite{Bern:2008qj, Bern:2010ue}, which plays a crucial role in our computation, refers to the conjecture that amplitudes or form factors in gauge theories can be organized into a representation such that
the kinematic numerators satisfy identities in one-to-one correspondence with color Jacobi identities. 
The duality puts strong constraints on the loop integrands, 
and when combined with the generalized unitarity method \cite{Bern:1994zx, Bern:1994cg, Britto:2004nc}, the construction of full-color loop integrands can be particularly efficient. Such a method has been successfully applied to get high-loop amplitudes in SYM \cite{Carrasco:2011mn, Bern:2012uf, Bern:2012cd,Bern:2012gh,Bern:2013uka,Bern:2013qca, Bern:2014sna, Johansson:2017bfl, Bern:2017ucb,Bern:2017yxu, Kalin:2018thp,Bern:2018jmv,Herrmann:2018dja} and pure YM \cite{Boels:2013bi, Bern:2013yya, Bern:2015ooa, Mogull:2015adi}, and also to form factors in ${\cal N}=4$ SYM \cite{Boels:2012ew,Yang:2016ear,Lin:2020dyj,Lin:2021kht,Lin:2021qol}.
While being proved at tree level \cite{BjerrumBohr:2009rd, Stieberger:2009hq, Feng:2010my},
the CK duality is still a conjecture at general loop orders and 
is not clear to what extent it applies. 
Actually, the constructions of the five-loop four-point amplitude in ${\cal N}=4$ SYM \cite{Bern:2017yxu, Bern:2017ucb} and two-loop five-point amplitude in pure YM \cite{Mogull:2015adi} have shown that the duality can be hard to realize.
On the other hand, the construction of three-loop form factors in \cite{Lin:2021kht,Lin:2021qol} suggests an alternative picture:  for form factors, the CK-dual solution space can be much larger, 
indicating that it is easier to achieve the duality.
In this paper, we extend the construction to the more non-trivial four-loop case. 
We find that the CK-dual solution space grows as the number of loops increases, and at four loops it contains 133 free parameters.

Given the multi-loop form factor integrands, we further study the hidden dual conformal symmetry in the  large-$N_c$ limit.
Similar to the aforementioned amplitudes /Wilson-loop duality, a duality between planar form factors and periodic Wilson lines was also proposed \cite{Alday:2007he,Maldacena:2010kp,Brandhuber:2010ad, Brandhuber:2011tv}. 
New features appear, however, due to the insertion of a color-singlet operator. For example, at one loop, the dual Wilson line computation requires a special truncation to Feynman diagrams within one period to match the form factor results \cite{Brandhuber:2010ad}. Moreover, the planar form factors involve integrals of non-planar topologies beyond one loop. Because of these, the form-factor/Wilson-line duality, as well as the dual conformal symmetry for form factors, are so far only understood at the one-loop level \cite{Brandhuber:2010ad, Bianchi:2018rrj}.
In this paper, we show for the first time that the planar form factor has a dual conformal symmetry in the lightlike limit $q^2\rightarrow0$, which we check explicitly up to four loops.

Before going into details, let us first give the final four-loop integrand result as follows:
\begin{equation}\label{eq:fullcolor-4loop}
 	\itbf{F}_{3}^{(4)} =  \sum_{\sigma_3} \sum_{i=1}^{229} \int \prod_{j=1}^{4} d^D \ell_j {1\over S_i} \,\sigma_3 \cdot {{\cal F}_{ 3}^{(0)}  C_i \, N_i \over \prod_{\alpha_i} P^2_{\alpha_i}} \,,
\end{equation}
where there are 229 cubic graphs,  $S_i$ are symmetry factors, $\sigma_3$ are the permutation operators exchanging external momenta and color indices, $C_i$ are color factors and $N_i$ are kinematic numerators. 
To get this, we employ a two-step strategy: (\Romannum{1}) to write down an ansatz by imposing the color-kinematics duality,
and (\Romannum{2}) to solve for the ansatz using diagrammatic symmetries and generalized unitarity cuts \cite{Bern:1994zx, Bern:1994cg, Britto:2004nc}. 
Our construction will follow closely the three-loop computation in \cite{Lin:2021qol}.
Reader is also referred to \cite{Bern:2012uf, Carrasco:2015iwa, Yang:2019vag} for further details.

\section{Integrand Construction (I): CK-dual Ansatz}
\label{sec:ansatz}

\begin{figure}[t]
\centering
\includegraphics[clip,scale=0.30]{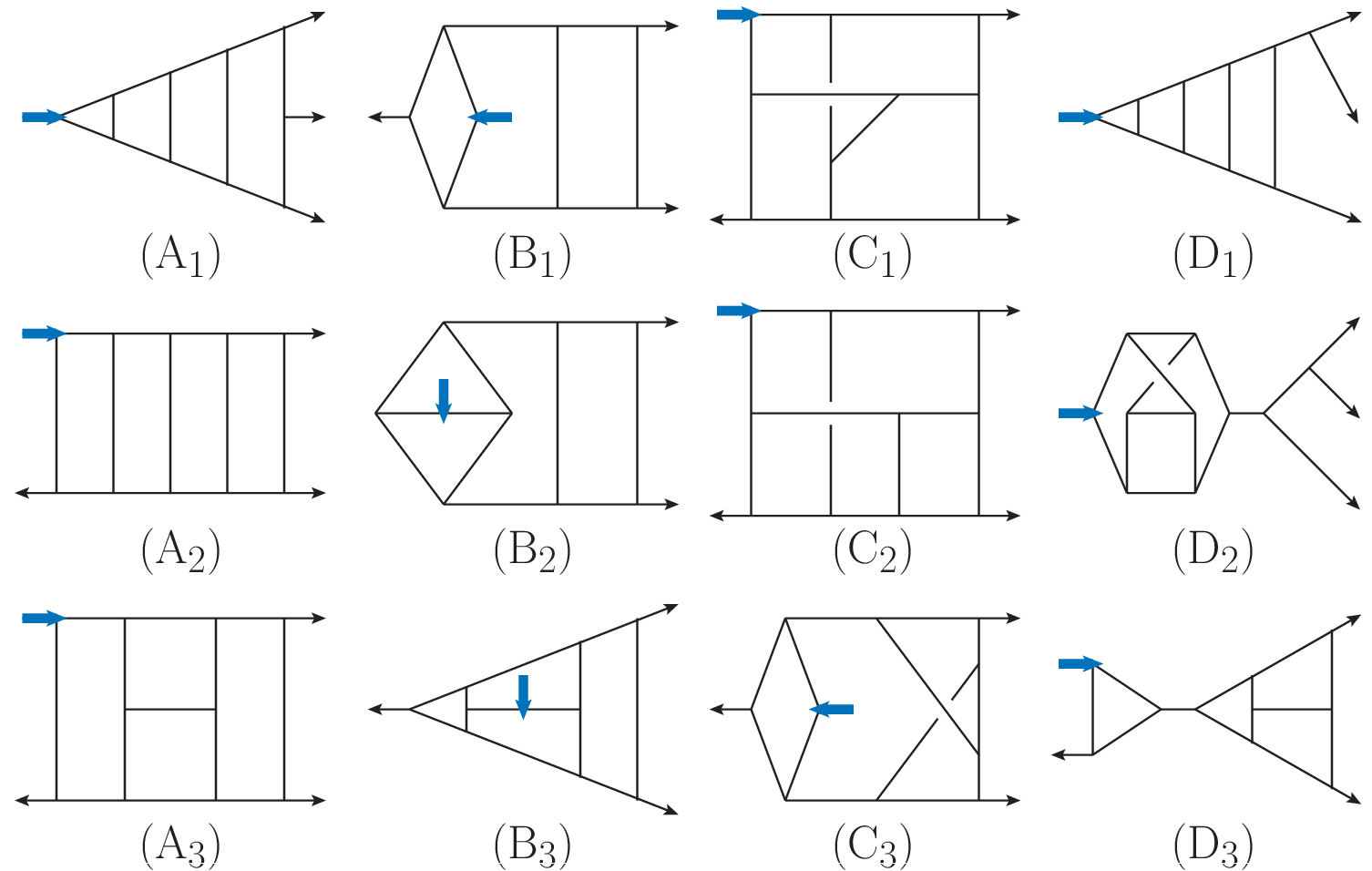}
\caption{Selected four-loop diagrams from the 229 topologies. }
\label{fig:phi2tops}
\end{figure}

\noindent
To begin with, we generate all trivalent diagrams $\Gamma_i$ with one operator $q$-leg and three on-shell legs. 
As observed in \cite{Bern:2012uf, Boels:2012ew,Yang:2016ear,Lin:2021kht}, for $\mathcal{N}=4$ SYM, it is reasonable to exclude diagrams with tadpole, bubble and triangle sub-graphs, unless the triangle is connected to the $q$-leg. 
Under these criteria, there are 229 trivalent topologies to consider. 
Selected examples are shown in Figure~\ref{fig:phi2tops}: 
the first column $(\text{A}_\text{i})$ are planar diagrams; 
the second column $(\text{B}_\text{i})$ includes diagrams defined as \textsl{$q$-interior planar} in the sense that after removing the color-singlet $q$-leg, the graphs are planar (they survive in the large-$N_c$ planar limit);
the third column $(\text{C}_\text{i})$ involves some intrinsic non-planar diagrams; 
some special one-particle-reducible graphs are shown in the last column as $(\text{D}_\text{i})$.

\begin{figure}[b]
\centering
\includegraphics[clip,scale=0.33]{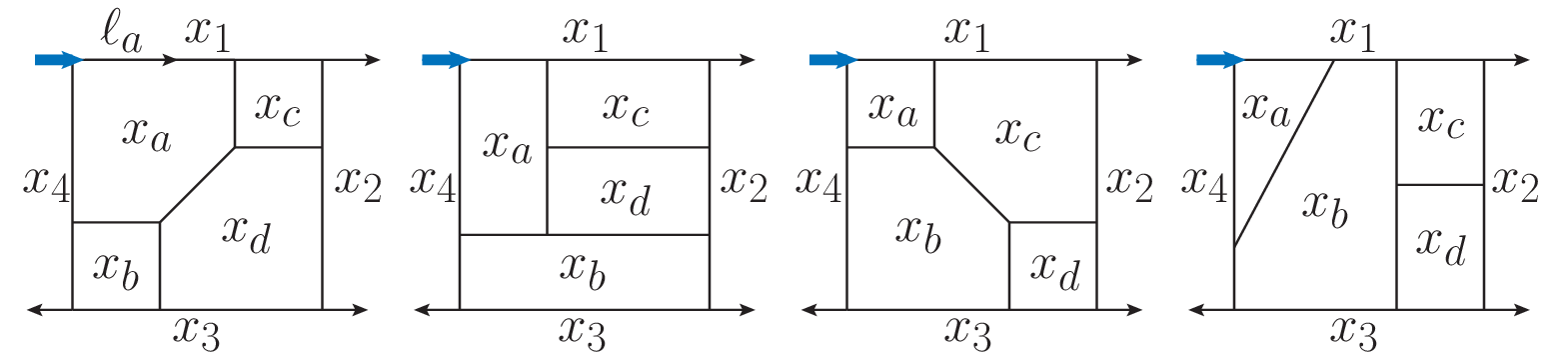}
\caption{Four planar master graphs. }
\label{fig:phi2master}
\end{figure}

The color factor $C_i$ and propagators $P_{\alpha_i}^2$  in \eqref{eq:fullcolor-4loop} can be read out from the corresponding trivalent diagram, whereas the genuine physical information is encoded in the kinematic numerators $N_i$. 
As mentioned above, the CK duality imposes strong constraints on numerators:
\begin{equation}\label{eq:dualJacobi}
	C_k = C_i + C_j  \ \  \Rightarrow \ \  N_k = N_i + N_j \,.
\end{equation}
We refer to the RHS as the dual Jacobi relations which interlock different kinematic numerators. 
With the help of dual Jacobi relations, we can use the kinematic numerators of a very small subset of graphs, \emph{i.e.}~the \textsl{master graphs}, to generate the numerators of all other diagrams.  
Practically, it is convenient to pick planar masters, and a minimal set containing four planar masters is given in Figure~\ref{fig:phi2master}.

Then we construct ansatzes for master numerators. 
It is convenient to parametrize the momenta using zone variables \cite{Drummond:2006rz}. For example, in the first diagram of Figure~\ref{fig:phi2master}, $\ell_a=x_{1}-x_{a}\equiv x_{1a}$, and the ansatz of the numerator $N_1^m $ is a polynomial of proper distances $x_{ij}^2$.
Moreover, since the form factor is half-BPS and has good UV behavior, we can impose the following power-counting constraints on the ansatz: 
a one-loop $n $-point sub-graph carries no more than $n-4 $ powers of the corresponding loop momentum \cite{Bern:2012uf}, with an exception that if the sub-graph is a one-loop form factor, the maximal power is $n-3$ \cite{Boels:2012ew}, see also \cite{Lin:2021qol} for details. 
As an example, for the first master in Figure~\ref{fig:phi2master}, monomials like $(x_{ij}^2)^2(x_{ai}^2)^2(x_{dj}^2)$ and $(x_{ij}^2)^3(x_{ad}^2)(x_{aj}^2)$, $i,j=1,2,3,4$, should be considered,  while $x_{b}$ and $x_{c}$ are not allowed to appear. 

In practice, we can simplify the ansatz construction by applying the symmetry as well as the maximal-cut properties of the master graphs. This can be done by starting from the rung-rule numerators \cite{Bern:1998ug,Bern:2010tq} and then adding terms proportional to propagators respecting the graph symmetries. 
For instance, the rung-rule numerator for the first master $N_1^{\rm m}$ can be obtained as 
\begin{equation}
\begin{aligned}
	N_1^{\rm m}|_{\text{rr}}=&x_{13}^2x_{24}^2 x_{a2}^2 x_{a3}^2(x_{d1}^2-x_{14}^2/2)- (x_{13}^2)^2x_{24}^2x_{a2}^2x_{d4}^2\\
	&-  x_{13}^2(x_{14}^2-x_{13}^2)(x_{a2}^2)^2x_{d4}^2+ (1\leftrightarrow4)\&(2\leftrightarrow3) ,
\end{aligned}
\end{equation}
which captures the maximal cut of the diagram. To further complement the ansatz, contributions involving propagators like $x_{d2}^2$ and $x_{d3}^2$ can be added in a symmetry-preserving way, such as 
 \begin{equation}
	N_1^{\rm m}=N_1^{\rm m}|_{\text{rr}}+\alpha_1 x_{13}^2 x_{24}^2 \big[ (x_{a2}^2)^2x_{d3}^2+(x_{a3}^2)^2x_{d2}^2 \big]+ \cdots.
\end{equation}
In the end, an ansatz of the CK-dual integrand with 1433 parameters for $\itbf{F}^{(4)}_3$ is obtained, and the four master numerators contain 257, 562, 479, and 135 parameters respectively.

\section{Integrand Construction (II): Constraints and Solution}
\label{sec:fixansatz}

\noindent
Given the ansatz, we now apply various symmetry and unitarity constraints to fix the parameters. 

First, similar to the master numerators, we require any other numerator $N_i$ to share the symmetry of the corresponding diagram $\Gamma_i$ and also reproduce the maximal cut. 
These conditions deal with only one numerator at a time and are practically very convenient to solve. Nicely, they put significant restrictions on the ansatz, reducing the number of parameters to 246.

\begin{figure}[t]
\centering
\includegraphics[scale=0.40]{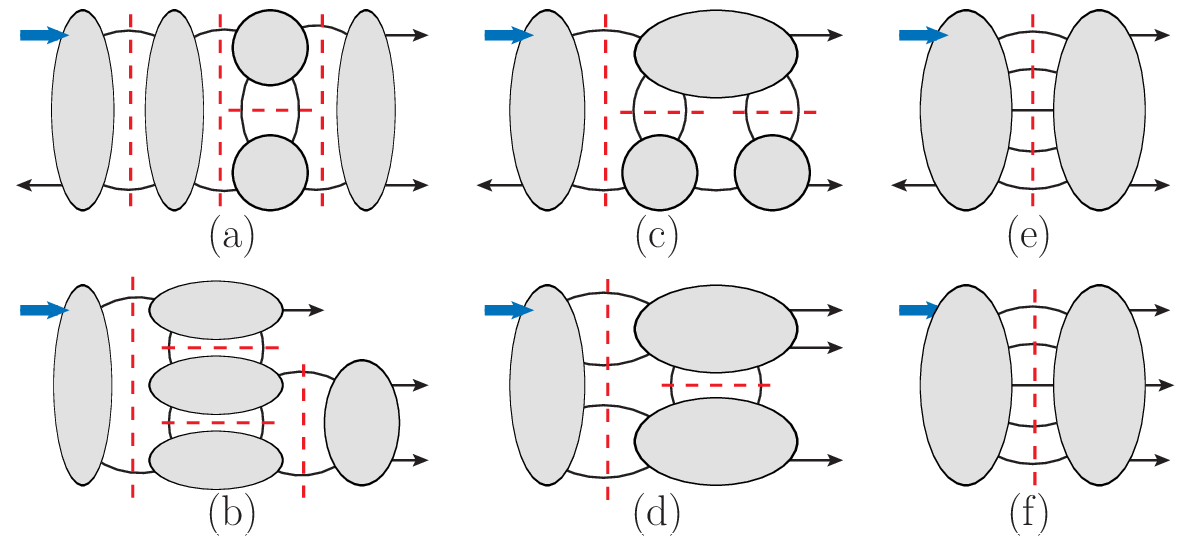}
\caption{Sample unitarity cuts. }
\label{fig:variouscuts}
\end{figure}

Next, we require the ansatz integrand \eqref{eq:fullcolor-4loop} to match all generalized unitarity cuts. 
Some typical cuts are shown in Figure~\ref{fig:variouscuts}. 
Cuts (a) and (b) are relatively simple octuple cuts, cutting the four-loop form factor into five tree blocks \footnote{If we are considering four-loop MHV form factors, all these blocks are actually simple MHV blocks.}. 
Such octuple cuts can be first conducted, which will eliminate 94 parameters.
Then we apply the septuple cuts and the sextuple cuts, such as Figure~\ref{fig:variouscuts}(c) and (d). These two types of cuts fix further 19 parameters. 
The most complicated cuts are the quintuple cuts like Figure~\ref{fig:variouscuts}(e) and (f).  
For instance, the cut (e) involves over a thousand cut-diagrams coming from the ansatz, of which the sum should reproduce the highly non-trivial tree product $\sum_{\rm helicity} \mathcal{F}^{(0)}_{5} \mathcal{A}^{(0)}_{8}$. We find that the quintuple cuts provide no more constraints on parameters. 
We emphasize that we have checked both planar and non-planar cuts, and details for performing cuts can be found in \cite{Lin:2021qol}. 
After all these cuts, we end up with a solution with 133 parameters and all dual Jacobi relations manifestly satisfied.
Thus, we obtain the CK-dual four-loop physical integrand given in \eqref{eq:fullcolor-4loop} with 133 parameters.

Finally, as an important check, the final form factor result must be independent of these free parameters. We check that all the free parameters manifestly drop out after performing a simplification of the integrand sketched as follows.
We first simplify the color factors by expressing the trivalent color factors $C_i$ in terms of $N_c$ and ${\tilde f}^{a_1a_2a_3}= {\rm tr}(T^{a_1}T^{a_2}T^{a_3}) -{\rm tr}(T^{a_1}T^{a_3}T^{a_2})$, giving both $N_c$-leading and $N_c$-subleading contributions as \footnote{
Let us give some details about \eqref{eq:simplified4loop}. There are 76 topologies in \eqref{eq:fullcolor-4loop} contributing to $ {\cal I}_{\rm pl}^{(4)}$, for example $(\rm{A_i})$ and $(\rm{B_i})$ in Figure~\ref{fig:phi2tops}; while
138 topologies contribute to $ {\cal I}_{\rm np}^{(4)}$, including $(\rm{C_i})$ in Figure~\ref{fig:phi2tops}.
Also, 28 topologies contribute to both $ {\cal I}_{\rm pl}^{(4)}$ and $ {\cal I}_{\rm np}^{(4)}$, such as  ($\rm{A_3}$) and ($\rm{B_3}$).
Note that 43 topologies have zero color factors and do not contribute to \eqref{eq:simplified4loop}, such as  ($\rm{D_2}$), but they are necessary to fulfill the CK duality.}
\begin{align}\label{eq:simplified4loop}
\itbf{F}_{3}^{(4)} & = {\cal F}_{3}^{(0)} {\tilde f}^{a_1a_2a_3} \big( N_c^4  \int  {\cal I}^{(4)}_{\rm pl} + N_c^2  \int {\cal I}_{\rm np}^{(4)} \big) \,.
\end{align} 
Next we manipulate ${\cal I}_{\rm pl}^{(4)}$ and ${\cal I}_{\rm np}^{(4)}$ respectively by expanding the integrands on a set of basis, following the procedure described in detail in \cite{Lin:2021qol}.
After this simplification, we achieve an equivalent but simplified integrand with all parameters eliminated. 

For the reader's convenience, we provide Supplemental Information on the detailed results:
the solution of four master numerators with 133 free parameters (as well as a particular solution with all the parameters assigned certain rational numbers), and
the other numerators $N_i $ as linear combinations of master numerators, together with the symmetry factors $S_i $, the color factors $C_i $, and the propagator lists ${P_{\alpha_i}} $ of all trivalent topologies in the form of \eqref{eq:fullcolor-4loop}.

\section{$q^2\rightarrow0$ limit and dual conformal symmetry}
\label{sec:solutionandcheck}

\noindent
Given the loop form factor integrands, we would like to explore the dual conformal symmetry. 
Rather than considering the full-color result, we will focus on the following limits.

(1) We take the planar large-$N_c$ limit. Compared with amplitudes, the form factor in the large-$N_c$ limit is much more involved: due to the existence of a color-singlet operator, there are indispensable contributions from $q$-interior graphs (as $(\text{B}_\text{i})$ in Figure~\ref{fig:phi2tops}), which are topologically similar to non-planar diagrams for amplitudes. These $q$-interior diagrams start to appear at two loops, and there are dozens more at four loops.  

\begin{figure}[t]
\centering
\includegraphics[scale=0.45]{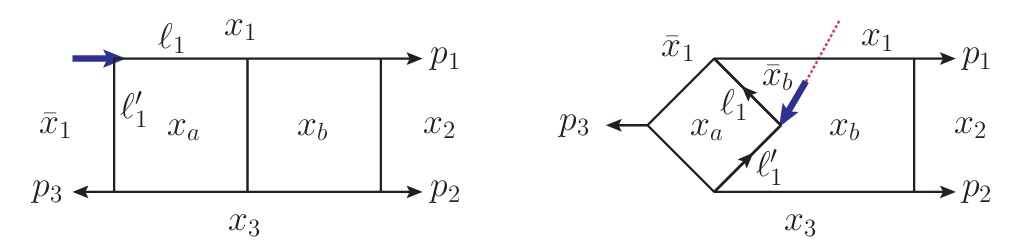}
\caption{Two-loop topologies contributing to the $q^2\rightarrow0$ limit.}
\label{fig:2loopDDCI}
\end{figure}

(2) We take the lightlike limit of $q$, namely $q^2 \rightarrow0$. This should be distinguished from the simpler soft limit $q=\sum_{i=1}^{3} p_i\rightarrow0$. In the latter case, one has $s_{12}=s_{23}=s_{13}=0$ and the integrand is trivially zero, since every numerator in the integrand contains at least two $s_{ij}$ factors (this applies also to the lower-loop cases).
In the lightlike limit, on the other hand, the form factor is simplified but still has a non-trivial integrand. 
For example, at the second loop order, two trivalent topologies survive in the $q^2\rightarrow 0$ limit as shown in Figure~\ref{fig:2loopDDCI}, where the second one is a crossed-box two-loop integral.

\begin{figure}[t]
\centering
\includegraphics[scale=0.4]{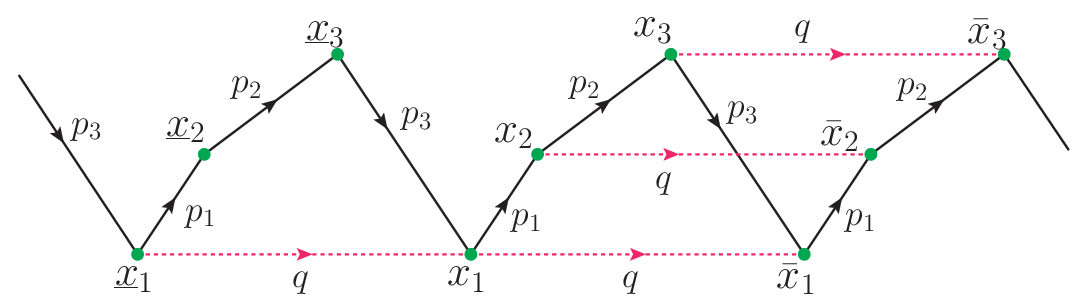}
\caption{Dual momentum space for the form factor.}
\label{fig:dualspace}
\end{figure}

Moreover, to study the dual conformal symmetry, we need to use the dual coordinates in momentum space, which can be represented by a periodic Wilson line configuration \cite{Alday:2007he,Maldacena:2010kp,Brandhuber:2010ad}.
For the three-point form factor, the (periodic) dual coordinates are defined as in Figure~\ref{fig:dualspace}:
\begin{equation}
x_i-x_{i+1} = p_i\,, \quad  \underline{x}_i - x_i = x_i - \bar{x}_i = q \,,
\end{equation}
where $q$ is the period of the periodic Wilson lines. 
Note that the previously mentioned zone variable $x_{4}$ in Figure~\ref{fig:phi2master} is labeled as $\bar{x}_1$ here with $x_{14}=x_{1\bar{1}}=q$ \footnote{We mention that to describe $N_c$-leading form factors, it is sufficient to introduce only $x_{1,2,3}$, $\bar{x}_{1}$ and the loop dual variables $x_{a,b,\ldots}$. When considering a general conformal boost $b^{\mu}$, these dual variables can simultaneously satisfy the transformation rule \eqref{eq:ddcitransformation}. If adding more $\bar{x}$, in general one needs to be careful about whether \eqref{eq:ddcitransformation} still holds for the new $\bar{x}$.}.
A special conformal transformation acting on the dual coordinates can be defined as
\begin{equation}\label{eq:ddcitransformation}
    \delta_{b}x_i^{\mu}=\frac{1}{2}x_i^{2}b^{\mu}-(x_i\cdot b)x_i^{\mu} \,,
\end{equation}
where $b^{\mu}$ is a conformal boost vector.

Now we are ready to state the dual conformal symmetry precisely: the four-dimensional planar form factor integrand in the $q$-lightlike limit, denoted as 
\begin{equation}
\mathbb{I}^{(\ell)}\equiv\mathcal{I}^{(\ell)}|_{\textrm{leading-}N_c \, \& \, q^2=0\, \& \, D=4}  \,,
\end{equation}
satisfies the directional dual conformal invariance (DDCI):
\begin{equation}
\label{eq:DDCI}
\delta_{q} \mathbb{I}^{(\ell)} = 0 \,,
\end{equation}
where $\delta_q$ is defined as in \eqref{eq:ddcitransformation} with $b^\mu \propto q^\mu$.
A similar DDCI symmetry was also considered for the non-planar four- and five-point amplitudes in \cite{Bern:2018oao, Chicherin:2018wes}.
It is worth mentioning that, for a general off-shell $q$, the special conformal transformation $\delta_q$ will break the periodic structure for the Wilson line; however,
$\delta_q$ with $q$ lightlike preserves the periodic configuration, with $\delta_{q}q=0$,  suggesting that $\delta_{q}$ is an exact symmetry for the form factor integrand. 

To give more details, the integrand $\mathbb{I}^{(\ell)}$ contains the integration measure and a rational function $R$ of proper distances $x_{ij}^2$ as
\begin{equation}\label{eq:specialintegrand}
\mathbb{I}^{(\ell)}= (\prod_a d^4 x_{a}) \times R(x_{ij}^2) \,,
\end{equation}
where the measure and the proper distance transform as
\begin{align}
\label{eq:DDCImeasure}
\delta_{q}\left(d^{4} x_{a}\right) &=-4 \left(q \cdot x_{a}\right) d^{4} x_{a}\,, \\
\label{eq:DDCIprop}
    \delta_{q}x_{ij}^2 &=-q\cdot(x_i+x_j)x_{ij}^2\,.
\end{align}
Note that the proper distance can include $\bar{x}_i$ or $\underline{x}_i$,
and importantly, the ``bar'' does not alter the weight
\begin{equation}
	 \delta_{q}x_{i\bar{j}}^2 =-q\cdot(x_i+x_j)x_{i\bar{j}}^2=-q\cdot(x_i+\bar{x}_j)x_{i\bar{j}}^2\,.
\end{equation}
In principle, it should be straightforward to verify the symmetry \eqref{eq:DDCI} using \eqref{eq:DDCImeasure}--\eqref{eq:DDCIprop} and Leibniz rules.
Some important remarks, however, are required. We will explain them with the help of some examples, in which the difference between the DDCI for form factors and an analog for the non-planar four- and five-point amplitudes in \cite{Bern:2018oao, Chicherin:2018wes} will also be highlighted.

\section{Remarks on the form factor DDCI}

\noindent
To begin with, the appearance of  $q$-interior topologies in form factors requires a proper definition of the dual coordinates. 
To do this,  practically one  can pull the $q$-leg to infinity, which introduces a cut in the plane.
The cut will divide several regions (or zones) into two pieces, and the periodic dual variables $x$ and $\bar{x}$ should be defined  on the two sides of the cut respectively.  

The two-loop form factor is a good example for us to work this out explicitly. As mentioned above, there are two contributing cubic diagrams given in Figure~\ref{fig:2loopDDCI}. It is trivial to define zone variables for the first planar diagram, while 
in the second diagram in Figure~\ref{fig:2loopDDCI}, we can introduce $x_1,\bar{x}_1$ on the two sides of the $q$-leg line so that all momenta and propagators can be expressed with the dual coordinates $\{x_1, x_2, x_3, \bar{x}_1, x_a, x_b\}$, for example
\begin{equation}
\ell_1^{\prime} = x_a - x_b \,,\quad \ell_1 = x_a - \bar{x}_b =  x_a-x_b+(x_1-{\bar x}_1) \,.  
\end{equation}
In general, for an $L$-loop form factor, one needs to introduce $(4+L)$ independent dual coordinates. 

Consider the two numerators for the two-loop diagrams in Figure~\ref{fig:2loopDDCI}. They are expressed in momentum variables as \cite{Brandhuber:2012vm}
\begin{equation}
	N_1(\ell_{i},p_{j})=N_2(\ell_{i},p_{j})= 2 s_{12} [ (\ell_{2}\cdot p_1) s_{13}- (\ell_{2}\cdot p_2) s_{23} ],
\end{equation}
and the propagators $D_{1,2}(\ell_{i},p_{j})$ can be also easily read out. The integrand $\mathbb{I}^{(2)}$ becomes
\begin{equation}\label{eq:specialintegrand2loop}
	\mathbb{I}^{(2)}=d^{4}\ell_{1}d^{4}\ell_{2}\bigg(\sum_{\sigma\in S_3}\sigma\cdot\frac{N_1}{D_1}+\sum_{\rho \in \mathbb{Z}_3}\rho\cdot\frac{N_2}{D_2}\bigg)\,,
\end{equation}
where $\sigma$ and $\rho$ acting on external lines $p_j$.

\begin{figure*}[t]
\includegraphics[width=0.9\textwidth]{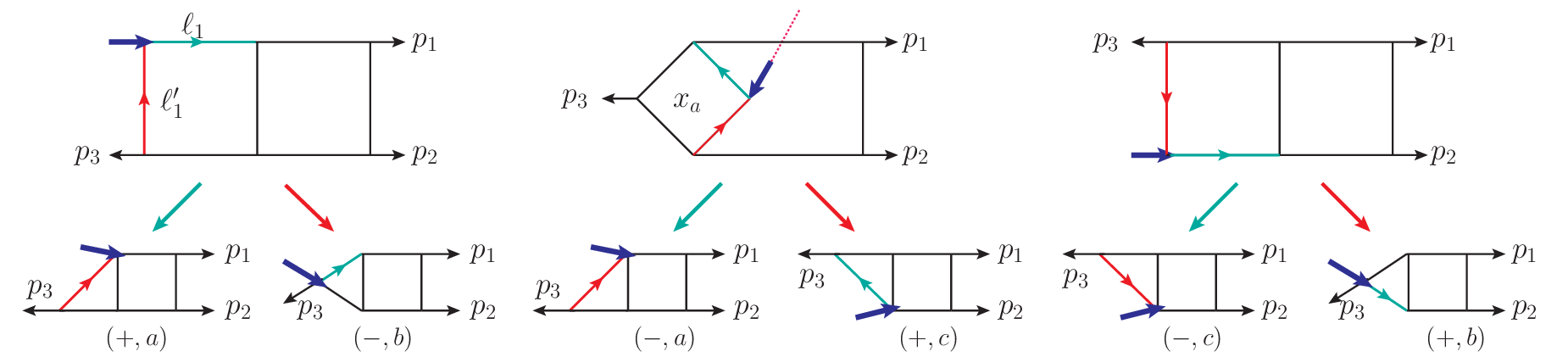}
\caption{The reduction of two-loop topologies after dual conformal transformation.}
\label{fig:cancel}
\end{figure*}

Now we can express these rational integrand ($N_{1,2}/D_{1,2}$) in terms of the dual coordinates for both diagrams, giving
\begin{equation}
\begin{aligned}
	N_{1}(x_{ij}^2)=&x_{13}^2\left(x_{13}^2(x_{a1}^2-x_{a2}^2)+x_{3\bar{1}}^2(x_{a1}^2-x_{a3}^2)\right)\,,\\
	N_{2}(x_{ij}^2)=&x_{13}^2\big(x_{13}^2(x_{a1}^2+x_{b2}^2-x_{a2}^2-x_{b1}^2)\\ & \qquad + x_{3\bar{1}}^2(x_{a1}^2+x_{b3}^2-x_{a3}^2-x_{b1}^2)  \big)\,,\\
	D_1(x_{ij}^2)&= x_{a1}^2 x_{a \bar{1}}^2 x_{a3}^2 x_{ab}^2 x_{b1}^2 x_{b2}^2 x_{b3}^2\,,\\
	D_2(x_{ij}^2)&= x_{a1}^2 x_{a3}^2 x_{a \bar{b}}^2 x_{ab}^2 x_{b1}^2 x_{b2}^2 x_{b3}^2\,,
\end{aligned}
\end{equation}
and we have the following result for $\delta_{q}$
\begin{equation}\label{eq:DDCIres2loop}
\begin{aligned}
	&\delta_{q} \ \frac{d^{4}x_{a}d^{4}x_{b} N_1}{D_1}=\frac{x_{a1}^2-x_{a\bar{1}}^2}{2} \frac{d^{4}x_{a}d^{4}x_{b} N_1}{D_1},\\
	& \delta_{q}\ \frac{d^{4}x_{a}d^{4}x_{b}N_2}{D_2}=\frac{x_{a\bar{b}}^2-x_{ab}^2}{2}\frac{d^{4}x_{a}d^{4}x_{b}{\color{blue}N_1}}{D_2}\,. 
\end{aligned}
\end{equation}
It is worthwhile taking a close look at the results: (1) The planar diagram is not conformal invariant, since we have one more power of $\ell_2$ in the numerator, in contrast with amplitudes in $\mathcal{N}=4$ SYM. (2) When it comes to the non-planar one, we see that the result in the second line of \eqref{eq:DDCIres2loop} involves not $N_2$ but $N_1$,  which  is highlighted by the color blue above. 
We can transfer \eqref{eq:DDCIres2loop} back to the momentum space 
\begin{equation}\label{eq:DDCIres2loop2}
\begin{aligned}
	\frac{x_{a1}^2-x_{a\bar{1}}^2}{2}\frac{{N_1(x_{ij}^2)}}{D_1(x_{ij}^2)}=\frac{\ell_1^2-(\ell_1^{\prime})^2}{2}\frac{{N_1(\ell_i,p_j)}}{D_1(\ell_i,p_j)}\,,\\
		\frac{x_{a\bar{b}}^2-x_{ab}^2}{2}\frac{{N_1(x_{ij}^2)}}{D_2(x_{ij}^2)}=\frac{(\ell_1^{\prime})^2-\ell_1^2}{2}\frac{{N_1(\ell_i,p_j)}}{D_2(\ell_i,p_j)}\,,
\end{aligned}
\end{equation}
where we have omitted the measure. 
We comment that the special conformal transformation $\delta_{q}$ is independent of the way how the $q$-leg is pulled to the infinity, for example  introducing  $\bar{x}_{a}$ rather than $\bar{x}_{b}$. 

In the end, one finds that $\delta_{q} \mathbb{I}^{(2)}=0$ after summing over all the permutations given in \eqref{eq:specialintegrand2loop}. To do this, we observe that from \eqref{eq:DDCIres2loop2},  contributions from both diagrams are proportional to propagators. Subtopologies can be generated by shrinking propagators, and the subtopologies originating from different parent ones cancel between each other, as illustrated in Figure~\ref{fig:cancel}.
We have used the fact that in \eqref{eq:DDCIres2loop2}, both the $\delta_q$ results involves two propagators ($\ell_1^2$ and $(\ell_1^{\prime})^2$) to shrink and gives two subtopologies. Also, these subtopologies (labeled by $a,b,c$) cancel pairwise, since the numerators are always proportional to the special $N_{1}$ factor but with different signs, which are also indicated in the parentheses at the bottom of  Figure~\ref{fig:cancel}.

We would like to stress several properties as follows.

First, in contrast to the amplitudes DDCI \cite{Bern:2018oao, Chicherin:2018wes}, the DDCI symmetry for form factors does not hold diagrammatically one-by-one. Only after summing over all the diagrams and their permutations can we show that the symmetry holds. 
In other words, the form factor integrand should be regarded as a whole object to satisfy the dual conformal symmetry. 
From the point of view of the integral expressions, this new feature (comparing to the amplitude case) is related to the fact that the loop momenta have higher powers for the sub-loops involving the $q$-leg, and this has its physical origin from the insertion of color-singlet operator for the form factor.

\begin{figure}[b]
\centering
\includegraphics[scale=0.33]{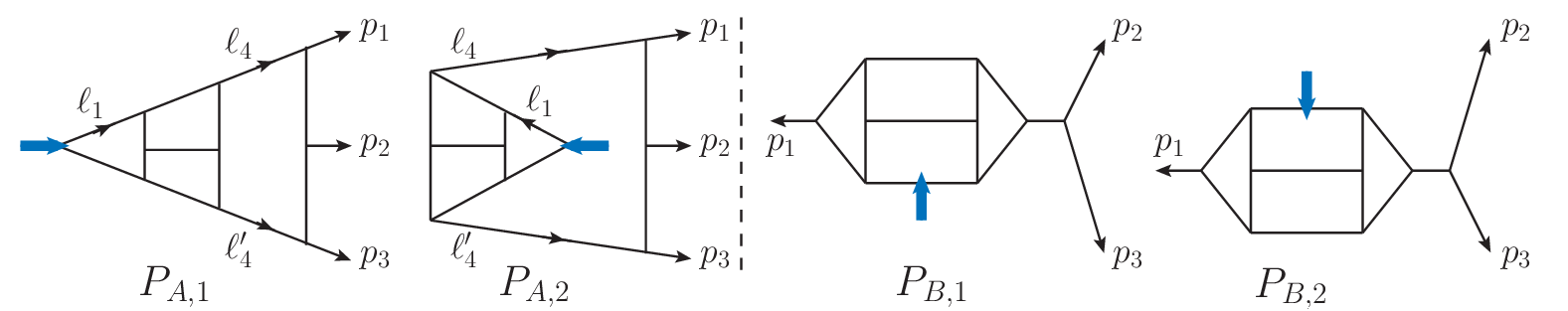}
\caption{Two examples of different planar projections for the same topology.}
\label{fig:triangleQleg}
\end{figure}

Another point worth mentioning is that some topologies require extra attention. These are topologies that have different planar projections, \emph{i.e.}, there are multiple ways of drawing a given topology on a two-dimensional plane without lines crossing\footnote{Finding all planar projections of a topology is equivalent to drawing all possible planar cubic diagrams (neglecting the $q$-leg)  and then inserting the operator $q$-leg. The location of the $q$-leg can be inside the planar diagram.}. 
Two such four-loop topologies are shown in Figure~\ref{fig:triangleQleg} \footnote{For one-, two- and three-loop integrands, after taking the limit $q^2\rightarrow0$, there exists ways of organizing results in terms of diagrams with unique planar projections. However, this can not be achieved for the  four-loop case.}.

Different planar projections will give different results after applying the special conformal transformation $\delta_{q}$. This is an intrinsicly new feature, irrelevant to how we parametrize the dual coordinates. Consider the first pair in Figure~\ref{fig:triangleQleg}. One can define dual coordinates, perform the $\delta_{q}$  transformation, and transfer the result back to the momentum space, just as in the previous two-loop case. 
However,  different planar projections will give very different results after the transformation. 
For example, we consider the first pair of diagram $P_{A,1}$ and $P_{A,2}$ in  Figure~\ref{fig:triangleQleg}. Using the momenta labeling given in Figure~\ref{fig:triangleQleg}, the numerators and propagators corresponding to these two diagrams are exactly the same. 
However, due to different planar projections, we have to introduce different dual coordinates:
for $P_{A,1}$ the dual coordinates are purely planar, while for $P_{A,2}$ it is inevitable to pull the interior $q$-leg out and introduce cuts and $\bar{x}_{i}$ coordinates. 
One can then evaluate the $\delta_q$ transformation for both cases in the dual space and transfer them back to the momentum space. 
Indeed, one finds that $\delta_{q}\mathbb{I}^{(4)}_{P_{A,1}}$ and  $\delta_{q}\mathbb{I}^{(4)}_{P_{A,2}}$ are different from each other, even though they are identical before the conformal transformation. 
Similarly, for $P_{B,1}$ and $P_{B,2}$ in Figure~\ref{fig:triangleQleg}, although they are isomorphic diagrammatically, they live in different dual projection space, and one has  to perform the $\delta_{q}$ calculation separately like the $P_{A,i}$ example. 
In conclusion, in order to get the correct DDCI transformation result, it is important to get the correct planar projections, and this can be determined by matching with the planar cuts \footnote{Practically one can generate all possible planar projections and distribute the integrand contribution equally among them.}.
Once again, the above subtlety is a new feature for form factors due to the operator insertion. 

As a final remark, we note that the free parameters introduced in the integrand construction appear to play no role in the DDCI discussion: 
the existence of these parameters does not break the symmetry, and the symmetry can not be used to further constrain these parameters \footnote{At the technical level, these statements mean that (1) one may perform a transformation on the integrand that depends on the parameters, and after simplifying the integrand as explained subsequent to Eq.~\eqref{eq:simplified4loop}, we find that not only do all the parameters cancel, but also that $\delta_{q} \mathbb{I}^{(\ell)} = 0$; and (2) one can try to constrain the numerator of each topology separately, for instance, to require that the master numerator satisfies DDCI independently, which is however not possible in general as can already be seen in the two-loop example.}.
This once again indicates that the form factor integrand possesses the dual conformal symmetry in an integrated manner.

\section{Conclusion and discussion}
\label{sec:discussion}

\noindent
In this paper we obtain the full-color four-loop integrand of the three-point form factor of the stress-tensor supermultiplet in ${\cal N}=4$ SYM. The color-kinematics duality has played a crucial role by providing a compact ansatz. 
At such a high loop order, it is a priori not clear if a CK-dual solution consistent with all unitarity cuts exists. 
Interestingly, we find the four-loop solution space contains a surprisingly large number of free parameters.
In Table~\ref{tab:fourloopsummary}, we give a summary of the CK-dual solutions up to four loops.
We can see that as the number of loops increases, the number of masters (thus the size of the ansatz) increases mildly, while the dimension of the CK-dual solution space grows significantly. This strongly suggests that the construction can be applied to the form factor at five and even higher loops.

Physically, these free parameters can be understood as duality-preserving deformations of the CK-dual integrands. 
For form factors, the existence of such deformations, and hence the exceptionally large solution space, is closely related to the generalized gauge transformation associated with the operator insertion, as discussed in the lower loop cases \cite{Lin:2021kht, Lin:2021qol}.
Moreover, this generalized gauge transformation is expected to have a connection with the double copy prescription for form factors for which the progress has been made recently at tree level \cite{Lin:2021pne}, and it would be interesting to explore this further.

 \begin{table}[t]
     \centering
     \caption{Statistics of CK-dual solutions for the three-point form factor up to four loops.  
     \label{tab:fourloopsummary}} 
     \vskip .1 cm
     \begin{tabular}{l|cccc}
     \hline \hline
     $L$ loops          & \, $L$=1 \,  & \, $L$=2 \, & \, $L$=3 \, & \, $L$=4 \, \\
      \hline 
     \# of cubic graphs & 2  &   6  & 29 & 229\\
     \# of planar masters & 1 &   2  &  2  & 4\\
     \# of free parameters & 1 & 4 & 24 & 133\\ \hline \hline
     \end{tabular}
 \end{table}

We also show that the planar form factor satisfies a directional dual conformal symmetry, namely,
the four-dimensional planar integrand in the limit of $q^2\rightarrow0$ satisfies precisely a directional dual conformal symmetry with the boost vector $b^{\mu}\propto q^{\mu}$. 
Such a hidden symmetry is non-trivial for form factors because non-planar diagrams are involved and are important in preserving the symmetry. It is also expected to be valid for general higher-point and higher-loop form factors of the stress-tensor supermultiplet, since the associated symmetry transformation preserves the periodic structure of the Wilson line configuration. Some two-loop four-point calculations under considerations support this symmetry, and more evidence is certainly expected.

For the \emph{integrated} planar form factors, the symmetry will be broken by infrared divergences, 
but the anomaly is expected to be captured by the anomalous conformal Ward identities \cite{Drummond:2007au}; 
and in this sense, the remainder functions of general $n$-point form factors are expected to have the DDCI property. A simple counting shows that the independent cross ratios appearing in the $n$-point remainder, in the $q^2\rightarrow 0$ limit, is $3n-9$. For the three-point case, the remainder will be pure transcendental numbers in the $q^2\rightarrow 0$ limit, and it is certainly beneficial to make a connection to the three-point planar form factors recently obtained up to remarkable eight loops \cite{Dixon:2020bbt,Dixon:2021tdw,Dixon:2022rse}, in which the $q^2\rightarrow 0$ limit corresponds to $u,v\rightarrow \infty$. For the higher-point cases, on the other hand, the remainders have non-trivial kinematic dependence in this limit, and the DDCI symmetry is expected to play a more non-trivial role.
It would be also interesting to explore the DDCI symmetry in the dual Wilson lines picture at both weak and strong coupling \cite{Alday:2007he,Maldacena:2010kp,Brandhuber:2010ad,Gao:2013dza}, as well as in the study of the operator product expansion (OPE) limit of the Wilson lines \cite{Sever:2020jjx, Sever:2021nsq}.
Since a similar DDCI has been observed for non-planar amplitudes \cite{Bern:2018oao, Chicherin:2018wes} (see also \cite{Ben-Israel:2018ckc}), one may hope that the study of planar form factors also helps to understand non-planar amplitudes. 

\vskip .2 cm

{\it Acknowledgments.}
It is a pleasure to thank Yuanhong Guo, Song He, Congkao Wen, Chen Xia and Mao Zeng for discussions.
This work is supported in part by the National Natural Science Foundation of China (Grants No.~11935013, 12175291, 11822508, 12047503), 
and by the CAS under Grants No.~YSBR-101 and XDPB15.
We also thank the support of the HPC Cluster of ITP-CAS.

\end{document}